# Eliminator Space and Its Application in Clutter Characterization and Suppression

Ziqian Wang
*China Electronics Corporation Nanjing Changjiang Electronics Group. Ltd*
Nanjing, China
ziqian86@mail.sitp.ac.cn

Pu Lu
*China Electronics Corporation Nanjing Changjiang Electronics Group. Ltd*
Nanjing, China
78079589@qq.com

***Abstract*—This paper presents a method for constructing slow-time-domain filters. Our approach establishes a deterministic signal space - referred to as the eliminator space - and applies subspace projection to reconstruct ground clutter components, which are then subtracted from the original data to enhance the detection probability of weak targets. Compared with conventional clutter filtering techniques, the proposed filter exhibits a flat passband response and an adjustable notch band. Additionally, the method entails low computational complexity in both space construction and filter implementation, while preserving the slow-time dimension of the signal. These characteristics improve radar target detection performance in low-altitude environments with strong ground clutter.**

***Keywords—Clutter suppression, Eliminator space, Subspace projection, Doppler filtering***

## I. Introduction

Enhancing the detection capability of "Low-Slow-Small (LSS)" targets tightly coupled with background clutter remains a fundamental challenge in radar engineering [1-5]. The standard solution to this challenge is pulse train detection in coherent radar systems, which leverages slow-time degrees of freedom and suppresses strong background clutter by constructing Doppler filters. The introduction of the slow-time dimension in pulse train processing enables radar signal processors to implement Doppler filtering, forming the foundation for detecting and tracking low-altitude targets with small radar cross sections (RCS) against static ground clutter backgrounds.

As a representative method in pulse train detection, moving target indicator (MTI) exhibits inherent limitations in detecting LSS targets in urban clutter environments, primarily due to its high passband ripple and significant loss for low-speed targets. Furthermore, for echoes with a limited slow-time dimension, MTI suffers from a dimensional mismatch between its output and input signals, which adversely affects the subsequent coherent integration after MTI filtering.

Modern radar information processors are capable of storing and processing all signals within a coherent processing interval (CPI) simultaneously, rather than conventional MTI filters that rely on pipeline processing [6-11]. For pulse train radar systems, the technique that performs matrix operations on a two-dimensional data frame with fast-time and slow-time dimensions and uses subspace projection to separate low-speed clutter from target echoes is clutter-noise separation. This is a powerful clutter filtering technique rooted in the idea of principal component analysis (PCA), a classical and widely adopted method in statistical data analysis, feature extraction, and data compression[12, 13].

Given a set of multi-dimensional observations, PCA identifies a subset of orthogonal eigenvectors that capture the maximum variance in the data with minimal redundancy. For radar echoes composed of clutter, targets, and noise, PCA performs eigen-decomposition on the data covariance matrix, yielding a small number of large eigenvalues and a large number of small eigenvalues. The eigenvectors corresponding to the small eigenvalues form the noise subspace, while those associated with the large eigenvalues constitute the clutter subspace. Following the PCA framework, an eigenvalue threshold to partition the eigenvectors could be defined. the eigenvectors corresponding to the first few largest eigenvalues are classified as clutter basis, and the remainder as noise basis vectors. Clutter-noise separation could be implemented by projecting the received echoes onto the noise subspace, or equivalently, subtracting the clutter components reconstructed from the clutter subspace, as shown in Fig.1. Conventional target detection algorithms can then be applied directly to the non-clutter components or signals in the noise subspace, while any residual echoes in the clutter subspace undergo further processing using techniques such as clutter maps (CMAP).

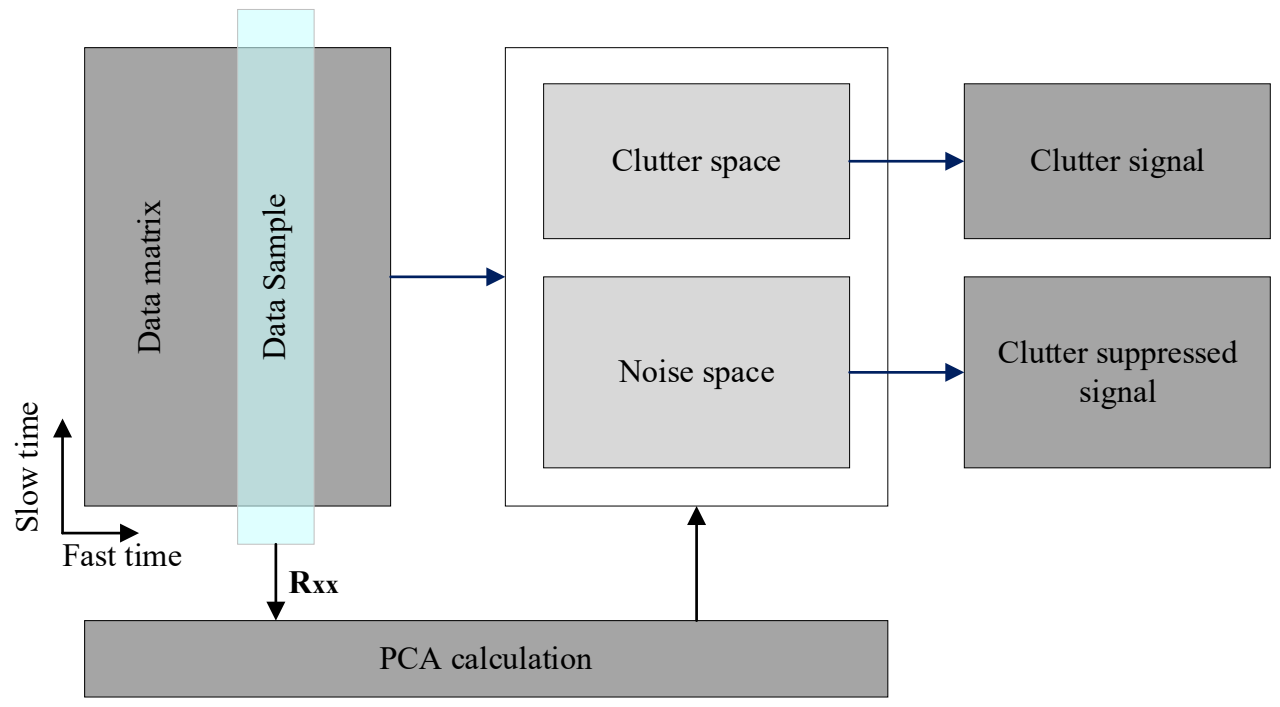

Fig. 1. Procedure of PCA-based clutter-noise separation

PCA can be used to estimate clutter spectrum parameters by collecting clutter samples, obtaining statistical information, and performing matrix decomposition to derive clutter eigenvectors. However, matrix eigen-decomposition involves high computational complexity. When combined with the enormous number of radar detection units, this renders the direct application of PCA impractical in radar signal processing systems with stringent low-latency requirements.

Strong ground clutter is primarily generated by diverse land surface features and buildings. Based on practical experience, we can establish a reasonable prior assumption: after compensating for the Doppler shift caused by the motion of the radar platform, ground clutter is essentially a low-frequency signal, and its average velocity is expected to be zero. Meanwhile, the spectral broadening of clutter Doppler spectra is caused by amplitude fluctuations from antenna beam scanning or the inherent statistical velocity distribution of clutter. Thus, even when the signal space corresponding to clutter echoes is obtained via PCA, its principal components exhibit remarkable consistency: the first eigenvector is approximately a direct current (DC) component, the second is nearly linear, and the third is typically quadratic. This distinct regularity and clear physical meaning of clutter characteristics motivate us to construct a deterministic signal space rather than conventional PCA approach.

## II. Methods

### A. Construction of Eliminator Space

We aim to construct an ideal clutter subspace composed of orthogonal eigenvectors through lightweight operations, thereby avoiding computationally cumbersome matrix inversion or eigen decomposition. This subspace can represent low-frequency signal components using the minimum number of eigenvectors, enabling efficient clutter suppression.

Double-pulse detection constitutes a fundamental form of pulse train detection. For this scenario, we first construct the most basic signal space. Departing from the conventional perspective that a double-pulse canceller is merely an FIR filter described by a difference function, we instead interpret it from the standpoint of signal space. It is evident that the double-pulse eliminator represented by the vector $\boldsymbol{\omega}_{1,2}=[1,-1]^T$, is an eigenvector of a two-dimension signal space. Consequently, it could be deduced that another eigenvector corresponds to an accumulator $\boldsymbol{\omega}_{2,2}=[1,1]^T$. The operation performed by the accumulator is equivalent to a second-order moving average filter. We correspondingly obtain the simplest signal space, which is spanned by the eliminator and the accumulator.

For the situation that there are N radar transmitted pulses during a CPI, an *N*-dimensional signal space denoted by $\mathbf{W_N}$ is required to perform subspace projection calculations in clutter-noise separation. This signal space can be constructed incrementally via a recurrence approach. The first N-1 eigenvectors $\boldsymbol{\omega}_{1,N}$, $\boldsymbol{\omega}_{2,N}$, …, $\boldsymbol{\omega}_{N-1,N}$ of $\mathbf{W_N}$ can be derived by applying a lifting operator $\mathbf{U}$ to all eigenvectors of $\mathbf{W_{N-1}}$:

$$\boldsymbol{\omega}_{i,N} = \mathbf{U}\boldsymbol{\omega}_{i,N-1}$$

Where the *n*-th element of $\boldsymbol{\omega}_{i,N}$ is deduced by:

$$\boldsymbol{\omega}_{i,N}(n) = n\boldsymbol{\omega}_{i,N-1}(n-1) + (N-n)\boldsymbol{\omega}_{i,N-1}(n)$$

Whereas the last eigenvector $\boldsymbol{\omega}_{N,N}$ is constructed by applying a generator $\mathbf{G}$ to the last eigenvector $\boldsymbol{\omega}_{N-1,N-1}$ of $\mathbf{W_{N-1}}$:

$$\boldsymbol{\omega}_{N,N} = \mathbf{G}\boldsymbol{\omega}_{N-1,N-1}$$

where

$$\boldsymbol{\omega}_{N,N}(n) = \begin{cases} \boldsymbol{\omega}_{N-1,N-1}(1) & , \quad n=1 \\ \boldsymbol{\omega}_{N-1,N-1}(n) - \boldsymbol{\omega}_{N-1,N-1}(n+1) & , \quad others \\ -\boldsymbol{\omega}_{N-1,N-1}(N-1) & , \quad n=N \end{cases}$$

Typically, after constructing an *N*-dimensional signal subspace, an additional normalization step is required, where the normalization operator is denoted by $\mathbf{O}$. Thus, the procedure for constructing an *N*-dimensional signal subspace is illustrated in Fig. 2a.

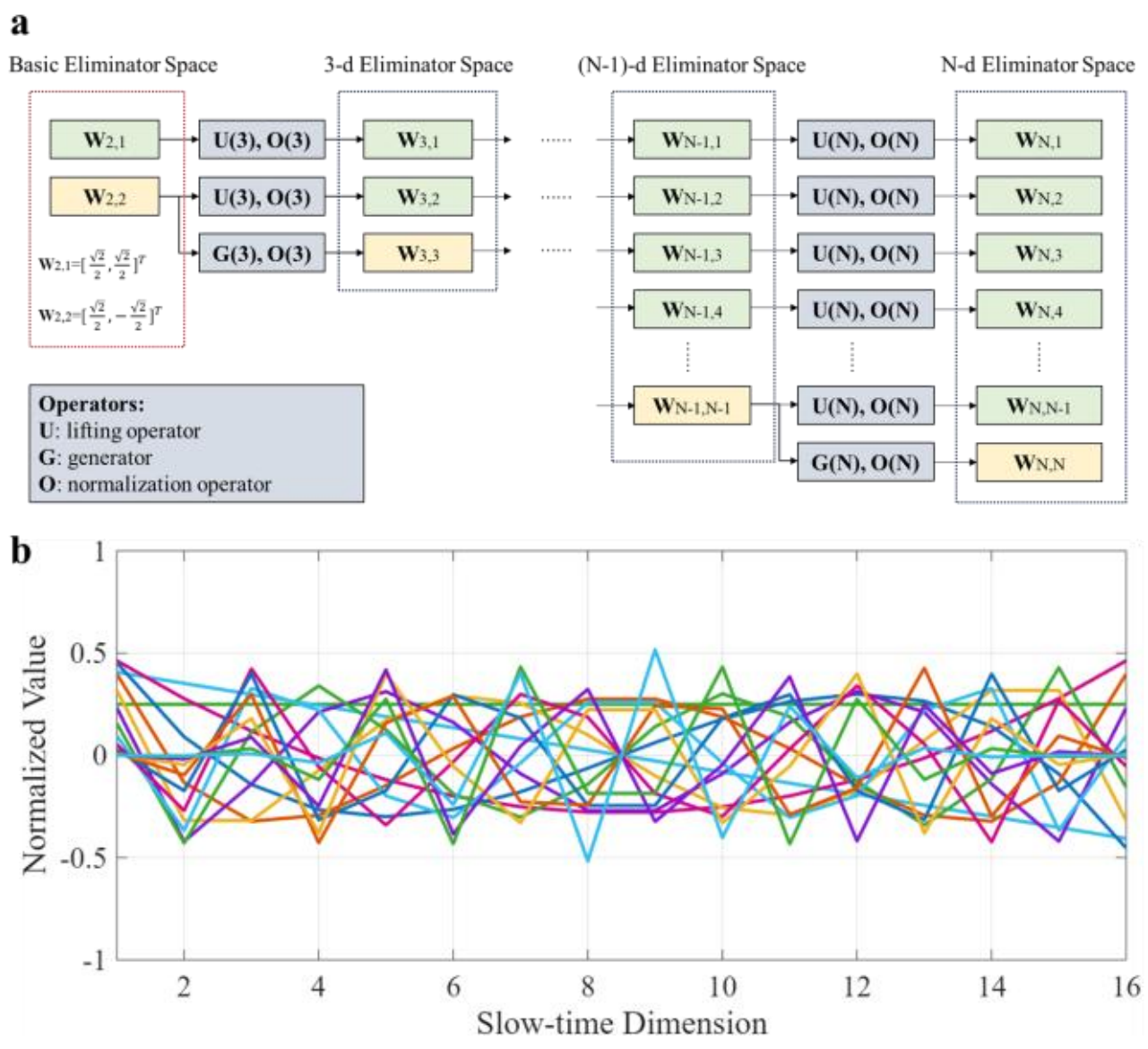

Fig. 2. a. procedure for constructing an N-dimensional eliminator space by a 2-D signal space and lifting operator $\mathbf{U}$, generator $\mathbf{G}$ and normalization operator $\mathbf{O}$. b. eigenvectors of a 16-D eliminator signal space.

This recurrence construction process of $\mathbf{W_N}$ supports rapid generation for arbitrary slow-time dimension, capable of accommodating both surveillance pulse-train waveforms with only several pulses and micro-Doppler detection waveforms with hundreds of pulses. The eigenvectors of an eliminator space $\mathbf{W_{16}}$ are demonstrated in Fig. 2b.

### B. Properties of Eliminator Space

We generate a set of orthonormal bases for an arbitrary number of pulses through simple arithmetic operations, thereby completing the construction of the eliminator space required for clutter-noise separation. This approach is compatible with the standard signal model of radar clutter, which states: ground clutter echoes consist of a low-frequency line spectrum component with a certain spectral broadening and an additive white Gaussian noise (AWGN) component. We hereby conclude that the eliminator space possesses the following two properties:

Firstly, eliminator space is a complete inner product space, i.e., a Hilbert space. Each eigenvector in an N-dimensional eliminator space is either symmetric or antisymmetric in the time domain. If the signal of interest is static ground clutter, all these eigenvectors are real-valued.

Secondly, the eliminator space is deterministic rather than sample-dependent. This constitutes a significant advantage for its rapid construction. It can describe the low-frequency components in the signal using the minimum number of eigenvectors, which enables the properties of this signal space to be comparable to or even superior to those of the clutter subspace obtained via PCA.

The structure of an *N*-dimensional eliminator has clear physical significance. Its eigenvectors are ordered such that the first eigenvector corresponds to the DC component of slow-time sequence, the second to the lowest-frequency AC component among all possible non-DC components, and so on to the last eigenvector.

In clutter suppression scenarios with multi-pulse detection, the last eigenvector is rarely used. Typically, only the first few low-frequency eigenvectors in clutter subspace participate in clutter suppression in engineering implementations. Nevertheless, the physical picture of the last eigenvector is crucial: it represents the signal component farthest from the DC component that can be extracted from a slow-time sequence. Notably, the last eigenvector of an $N$-dimensional EASS corresponds exactly to an $N$-pulse canceller.

### *C. Clutter-Noise Separation in Eliminator Spaces*

After beamforming, the radar echoes from pulse train detection are input to the radar signal processor in the form of a two-dimensional data matrix. Clutter-noise separation is performed in the eliminator space is analogous to that conducted in the clutter subspace formed by PCA. However, the entire procedure is simplified because clutter sampling and the subsequent calculation of the signal space based on the sampled data is not required

Prior to processing the beam signals, the parameters of the eliminator space are first determined. The required dimension of the eliminator space, denoted as $N_{ES}$, is equal to the number of pulses, noted by $N_P$, in the radar pulse train, and the boundary between the clutter subspace and the noise subspace is designed to meet the required clutter suppression performance. This boundary is established by setting the dimension of the clutter subspace, denoted as $N_C$. For example, to achieve a suppression level of more than -60 dB for a clutter spectrum with an average velocity of 0 and a standard deviation of 1 m/s, given radar parameters of PRI of 500 µs, 32 pulses per CPI, and a carrier frequency $f_c$ = 10 GHz, this requirement can be satisfied when $N_C$ = 4, as shown in Fig. 4a. The implementation of this step is deterministic. During radar signal processing, parameters of the signal space and the dimension of the clutter subspace are pre-stored in the signal processor.

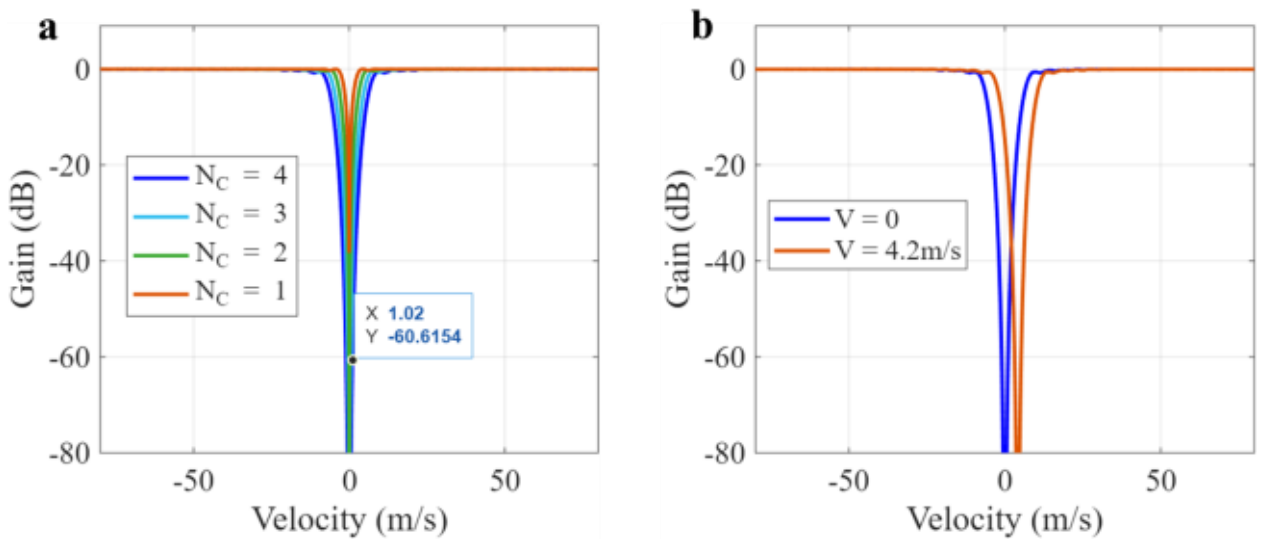

Fig. 3. a. Notch characteristics of eliminator subspace projection filters at different $N_C$. b. Effect of shifting the filter notch positions by rotation.

When performing clutter-noise separation, each slow-time sequence of the echo signal matrix $\mathbf{X_R}$ is first projected onto the clutter subspace to reconstruct the clutter signal $\mathbf{X_C}$. The reconstructed clutter signal is then subtracted from the slow-time sequence data, yielding the signal $\mathbf{X_N}$ in the noise subspace.

In this step, the slow-time dimension of the radar signal is preserved, which represents a significant advantage of clutter-noise separation approach over MTI. When the slow-time dimension is $N_P$, processing MTI through a classical three-pulse canceller will reduce the slow-time dimension of the output signal to $N_P$-2. In contrast, both the noise subspace signal $\mathbf{X_N}$ and the reconstructed clutter signal $\mathbf{X_C}$ output by clutter-noise separation maintain the same slow-time dimension as the input signal.

The notch depth of the clutter-noise separation filter can also be adjusted by selecting appropriate dimension of the clutter subspace, which enables the filter to adapt to clutter of different intensities, as shown in Fig. 3. Additionally, when dealing with moving clutter with non-zero valued velocity, which requires shifting the notch position to the desired Doppler frequency $f_d$ of the clutter suppression filter, this adjustment can be achieved by introducing a diagonal rotation matrix that consists solely of diagonal elements, as demonstrated in Fig. 3b.

We first validate the clutter suppression performance of the eliminator space on an S-band low-altitude warning radar. The key detection parameters of this radar are summarized in Table 1. A frame of radar beam data consists of 40 pulse echoes, as illustrated in Fig. 4a. Clutter is distributed over the entire radar range coverage.

TABLE I. DETECTION PARAMETERS OF AN S-BAND RADAR

| Carrier frequency | Bandwidth | PRI | Polarization | Mode |
|---|---|---|---|---|
| 2.71GHz | 10MHz | 405µs | HH | Electronic scan |

Carrier frequency bandwidth PRI polarization Mode

We construct a 40-dimensional eliminator space, and select $N_C$ = 2 for clutter-noise separation. Since this radar operates in the electronic scanning mode, the clutter spectrum was not broadened by beam scanning. Thus, two clutter eigenvectors is sufficient to meet the requirements of clutter suppression. After projecting the radar echoes onto the clutter subspace, the reconstructed clutter signal and the corresponding orthogonal signal in the noise subspace are presented in Fig. 4b and Fig. 4c, respectively.

Subsequently, coherent integration processing is performed on the projection results in the noise subspace. Here, discrete Fourier transform (DFT) is adopted to construct a Doppler filter bank for different radial velocities. Since clutter is suppressed during clutter-noise separation, target echoes can be clearly observed in the range-Doppler two-dimensional data in Fig. 4d.

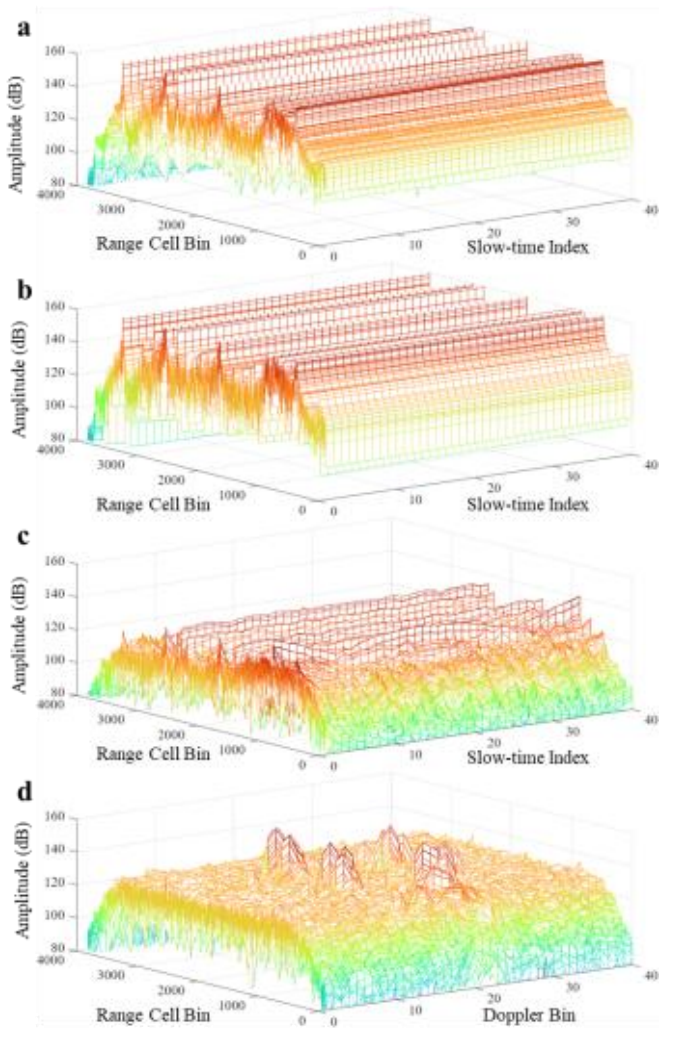

Fig. 4. Intensity distributions of data matrices in the S-band surveillance radar: a. beam data matrix, b. reconstructed clutter echo via clutter subspace projection, c. data matrix obtained from the noise subspace, and d. coherent integration results of clutter-suppressed echoes after MTD processing.

The performance of this clutter suppression technique is also verified on another X-band coastal warning radar. The detection parameters of this radar are listed in Table II. A frame of radar beam data, which contains high elevation and low elevation beams, consists of 128 pulse echoes. We construct a 128-dimensional eliminator space and select its first five eigenvectors to characterize clutter echoes. Unlike the S-band radar, this radar operates in a mechanical scanning mode and exhibits a broader clutter spectrum, thus more eigenvectors are required to characterize the clutter.

TABLE II. DETECTION PARAMETERS OF AN X-BAND RADAR

| **Carrier frequency** | **Bandwidth** | **PRI** | **Polarization** | **Mode** |
|---|---|---|---|---|
| 9.42GHz | 20MHz | 88.5μs | VV | mechanical scan |

We present the processing results of clutter-noise separation on the two beam datasets here. The two-dimensional echo data of the high-elevation beam is shown in Fig. 5a. After projecting the radar echoes onto the clutter subspace, the reconstructed clutter signal and the corresponding signal in the noise subspace are shown in Fig. 5b and Fig. 5c, respectively. Subsequently, MTD processing is applied to the projection results in the noise subspace using the same Doppler filter bank approach. As clutter is effectively suppressed during the clutter-noise separation process, target echoes become clearly visible in the range-Doppler two-dimensional data.

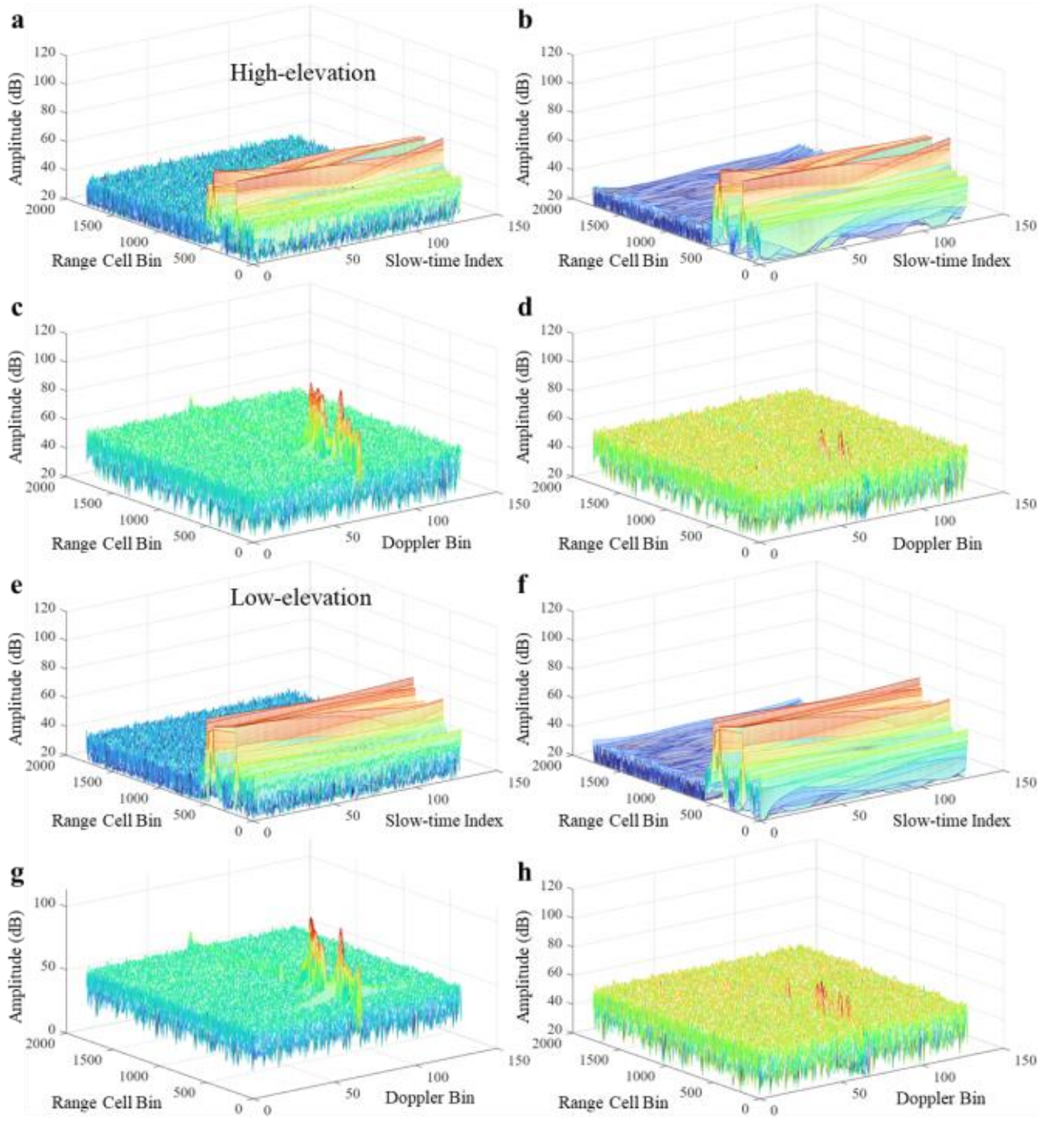

Fig. 5. Intensity distributions of data matrices from the X-band radar. a–d High-elevation beam: beam data matrix, reconstructed clutter echo, raw MTD results, and clutter-suppressed MTD results; e–h Low-elevation beam: the corresponding results in the same order.

The same processing procedure is applied to the low-elevation beam data (Fig. 5e), and the corresponding results are shown in Figs. 5f to 5h. Since the two beams share the same azimuth, their clutter distribution and target distribution exhibit high similarity. However, there are certain differences in intensity distribution, which is attributed to the elevation of different scatters. The clutter-noise separation technique with identical processing parameters achieves excellent clutter suppression performance for echoes in both the high-elevation and low-elevation beams.

## III. CONCLUSION

The eliminator space is proposed for low-speed small target detection in clutter environments. We introduce the concept of the eliminator space and a construction method, which can recursively span the signal space of the required dimension from the basic two-dimensional eliminator space (composed of an eliminator and an accumulator), and pre-store it in the radar signal processor in a deterministic form. The construction of the eliminator space does not depend on the selection of clutter samples, nor does it require extensive matrix calculation. From the eliminator space, we can define the clutter subspace and the noise subspace, which are then applied to the clutter-noise separation process. We validate the performance of clutter-noise separation in the eliminator space using echo data from two radars. The low-frequency clutter components are effectively suppressed, and the detection probability of low-speed targets is significantly improved.